\documentclass[a4paper,fleqn]{cas-dc}

\usepackage{soul,color}
\usepackage[numbers, sort&compress]{natbib}
\usepackage{lipsum}
\usepackage{booktabs}

\def\tsc#1{\csdef{#1}{\textsc{\lowercase{#1}}\xspace}}
\tsc{WGM}
\tsc{QE}
\tsc{EP}
\tsc{PMS}
\tsc{BEC}
\tsc{DE}

\newcommand{\KnR}{ \text{Kn}^{-1}_R }
\newcommand{\IKn}{ \text{Kn}^{-1}}
\newcommand{\mfp}{ \lambda_{\text{mfp}}}
\newcommand{\trento}{T\raisebox{-.5ex}{R}ENTo}

\begin{document}
\let\WriteBookmarks\relax
\def\floatpagepagefraction{1}
\def\textpagefraction{.001}
\shorttitle{Knudsen number and universal behavior of collective flows in conformal and non-conformal systems}
\shortauthors{V. Nugara et~al.}

\title [mode = title]{Knudsen number and universal behavior of collective flows in conformal and non-conformal systems} 



\author[1,2]{Vincenzo Nugara}[orcid=0009-0006-1939-8663]

\author[3]{Nicolas Borghini}[orcid=0000-0002-3906-3258]

\author[1,2]{Vincenzo Greco}[orcid=0000-0002-4088-0810]

\author[1,2]{Salvatore Plumari} [orcid=0000-0002-3101-8196]

\address[1]{Dipartimento di Fisica e Astronomia "E. Majorana", Università degli Studi di Catania, Via S. Sofia 64, 1-95125 Catania, Italy}
\address[2]{Laboratori Nazionali del Sud, INFN-LNS, Via S. Sofia 62, I-95123 Catania, Italy}
\address[2]{Fakult\"at f\"ur Physik, Universit\"at Bielefeld, D-33615 Bielefeld, Germany}

\begin{abstract}
We investigate the role of the Knudsen number (Kn) as a scaling parameter governing the emergence of collective behavior in relativistic heavy-ion collisions. Using the Relativistic Boltzmann Transport approach, we explore different initial conditions for both conformal (massless) and non-conformal (massive) systems with a constant specific shear viscosity $\eta/s$. Observables such as the time evolution of anisotropic flow coefficients collapse onto universal curves for fixed classes of Knudsen number, when using a scaled time variable accounting for the system size and the speed of sound $c_s$. More differential quantities, such as $v_n(p_T/\langle E_T\rangle)$, show a larger sensitivity to $c_s$. We also study events with fluctuating initial profiles from the \trento\ model, simulating collision systems from O-O to Pb-Pb at RHIC and LHC energies. Universal scaling at a given Kn value also holds in these event-by-event simulations, suggesting that the Knudsen number provides a unified criterion for classifying collectivity across different systems, including small systems where thermalisation may not be fully realised.

\end{abstract}

\maketitle

\section{Introduction}
Ultra-Relativistic Heavy-Ion Collisions (uRHICs) are a powerful probe to investigate how strongly interacting matter behave under extreme conditions. 
One of the key phenomena investigated in this field is the possible phase transition, suggested by experimental evidence, to a deconfined medium of quarks and gluons: the Quark-Gluon Plasma (QGP). A first-principle description of the formation and evolution of such medium by solving the underlying theory of QCD is still beyond our computational capabilities; therefore, in order to infer the fundamental properties and model the evolution of QGP, various effective dynamical models have been proposed. Both macroscopic hydrodynamics theories \cite{Teaney:2009qa, Gardim:2012yp, Romatschke:2017ejr, Florkowski:2017olj, Heinz:2013th, Denicol:2012cn} and mesoscopic kinetic approaches \cite{Ruggieri:2013bda, Ruggieri:2013ova,  Ruggieri:2015yea, Plumari:2015sia, Plumari:2019gwq, Xu:2004mz,Cassing:2009vt,Soloveva:2021quj, Kurkela:2018vqr} have been successful in describing and predicting the collective behaviour of this medium. It is commonly assumed that after a pre-equilibrium stage, necessary for the medium to reach at least a partial equilibration, an hydrodynamic phase sets in, until the extremely fast expansion cools down and rarefies the medium in $\gtrsim 10$ fm.  It is now clear, by comparing collective flows from nucleus-nucleus experiments with hydrodynamics simulations, that the produced medium behaves as an almost perfect fluid, exhibiting the smallest shear viscosity over entropy density ratio ever measured ($\eta/s\approx 0.1$). Surprisingly, experimental data have shown significant azimuthal anisotropies also in smaller systems, such as proton-nucleus ($pA$) and proton-proton ($pp$) collisions in high multiplicity events \cite{ALICE:2014dwt, ATLAS:2017hap, CMS:2017kcs, Grosse-Oetringhaus:2024bwr, Loizides:2016tew}. These systems are smaller (the volume is expected to be $\sim 10^{-2}$--$10^{-3}$ that of an $AA$ collision), live for a shorter time and probably never reach full thermalisation; nonetheless viscous hydrodynamics can still successfully reproduce experimental data. This raises  questions about the possible formation of droplets of QGP even in small collision systems and challenges the traditional regime of validity of hydrodynamics, which has been pushed beyond its conventional limits of applicability. Future runs with O-O, Ne-Ne or Ar-Ar collisions at LHC promise to shed light on the question, since these medium-size systems are expected to bridge the gap between $AA$ and $pA$ or $pp$, and thus to clarify the transition between the hydrodynamic and the particle-like regimes. 
From a theoretical perspective, the problem of hydrodynamisation has been related to the emergence of attractors. Studies performed in different frameworks (hydrodynamics, kinetic theory, classical Yang--Mills dynamics, AdS/CFT correspondence~\cite{Heller:2015dha, Strickland:2017kux, Chattopadhyay:2019jqj, Jaiswal:2019cju, Blaizot:2019scw, Heller:2020anv, Kurkela:2018vqr, Kurkela:2019set, Almaalol:2020rnu, Behtash:2017wqg, Blaizot:2017ucy, Strickland:2018ayk, Heller:2018qvh, Tanji:2017suk, Brewer:2022vkq, Ambrus:2021sjg, Berges:2013fga, Heller:2011ju, Du:2023bwi, Nugara:2023eku, Nugara:2024net}) have explored whether and how the reaching of the attractor, and therefore of a universal behaviour, can justify the success of hydrodynamics even far away from equilibrium \cite{Romatschke:2017vte}. 
According to the standard formulation of the hydrodynamic theory, its applicability is subjected to two conditions: the medium must be close to local thermal equilibrium and the microscopic and macroscopic scales of the system must be well separated. The ratio between these two scales, known as Knudsen Number (Kn), quantifies whether the microscopic degrees of freedom of the system are necessary for a macroscopic description, i.e. whether or not the system could be in principle modeled as a fluid. In this work we aim to use the Knudsen number to label different collision systems and give an effective description in terms of a universal behaviour.

Recently, in the context of the Relativistic Boltzmann Transport (RBT) approach \cite{Nugara:2024net}, it was suggested that the Knudsen number may act as the key scaling parameter in the massless kinetic theory. In this Letter, we investigate this idea in a detailed and systematic way, extending for the first time the model also to the non-conformal (massive) systems. In addition, we investigate the role of the speed of sound in the evolution of the observables and its impact in the onset of universality. Finally, we include in our model initial state fluctuations to simulate different collision systems (Pb-Pb, Au-Au, Ne-Ne, O-O) and study the presence of universality in terms of Knudsen number in this more realistic context. 

\section{Relativistic Boltzmann Transport and initial conditions}
\label{sec:RBT}
We employ the relativistic transport code developed over the past years to perform studies on the QGP dynamics, for a broad range of collision systems from RHIC to LHC energies 
\cite{Ferini:2008he, Plumari:2012ep, Scardina:2012mik, Scardina:2014gxa, Plumari:2015sia, Plumari:2015cfa, Scardina:2017ipo, Plumari:2019gwq, Plumari:2019hzp, Sun:2019gxg, Sambataro:2020pge, Gabbana:2019uqv, Nugara:2023eku, Nugara:2024net}.
In the present work, the hot QCD matter is described as a degenerate system of partons, which amounts to simulating a one-component fluid. We model it by means of an on-shell one-body
distribution function $f$, depending on space-time coordinates $x^{\mu}=\left(t,\mathbf{x}\right)$
and 4-momentum $p^{\mu}=(p^0,\mathbf{p})$, with $p^{0}=E_{\mathbf{p}}\equiv\sqrt{m^{2}+\mathbf{p}^{2}}$. The space-time evolution of $f$ is
governed by the following RBT equation
\begin{equation}
p^{\mu}\partial_{\mu}f(x,p)=C\left[f(x,p)\right]_{\mathbf{p}},
\label{eq:RBT}
\end{equation}
with $\partial_{\mu}\equiv\partial/\partial x^{\mu}$ denoting the gradient with respect to space-time coordinates $x^{\mu}$.
We implement only elastic $2\leftrightarrow 2$ collisions, which leads to a particle-number conserving system with a space-time dependent fugacity $\Gamma(x)\ne 1$.  The collision integral is therefore:
\begin{align*}
C\left[f\right]_{\mathbf{p}} =&
\intop d\Gamma_2 \intop d\Gamma_{1'} \intop d\Gamma_{2'}\left(f_{1'}f_{2'}-f_{1}f_{2}\right) \\ &\times \left|\mathcal{M}\right|^{2}\delta^{\left(4\right)}\left(p_{1}+p_{2}-p_{1'}-p_{2'}\right),
\end{align*}
where $f_i=f(p_i)$ and $d\Gamma=d^3p/((2\pi)^3 E_p)$. $\mathcal{M}$ denotes the transition amplitude for the elastic processes {$|{\cal M}|^2=16 \pi\,\mathfrak{s}^2 d\sigma/d\mathfrak t$} with {$\mathfrak{s}$} and  {$\mathfrak t$} the Mandelstam variables. In this work, we restrict ourselves to collisions with isotropic cross sections; however, in the regime of validity of hydrodynamics, the dynamical evolution is independent on the details of the differential cross section.
We determine locally in space and time the total cross section $\sigma_{\rm tot}$ in order to fix $\eta/s$.
An analytical relation between $\eta$, the temperature $T(x)$ and $\sigma_{\rm tot}$ is obtained via the Chapman--Enskog approximation at second order for massive particles:
\begin{equation}
\eta(x)=f(z)\frac{T(x)}{\sigma_{\rm tot}(x)};
\label{eq:sigma}
\end{equation}
see Ref. \cite{Plumari:2012ep} for details on $f(z)$ with $z(x)=m/T(x)$. In the limit $z\to0$, this function goes to $f(z)\to 1.258$, which differs from the 16th order result $1.267$ by less that $1\%$.
It has been shown in \citep{Gabbana:2019uqv} that this approach is able to describe the dynamical evolution in a wide range of $\eta/s$, even below $\eta/s=0.05$ which is smaller than the  conjectured minimal bound of $1/4\pi$ for the QGP \cite{Kovtun:2004de}. Notice that, in the non-conformal case, the system also has a bulk viscosity $\zeta$, which depends on mass, temperature and cross section. By choosing the local cross section to keep $\eta/s$ constant, we fix locally the bulk viscosity as well. However, since the system is particle-number conserving, the specific bulk viscosity is expected to be very small: from analytical approaches (see e.g.~\cite{Ambrus:2023qcl}), $\zeta/s$ is of order $10^{-4}$. We estimated its value in our numerical simulations by comparing them with viscous hydrodynamic results obtained with MUSIC \cite{Schenke:2010nt, Schenke:2010rr, Paquet:2015lta}: we observe a very good agreement between the two approaches if we set $\zeta/s\sim 10^{-4}$ in MUSIC. 
Such a small $\zeta/s$ value is admittedly not realistic, but the focus of the present study is mainly to investigate the role of a non-conformal equation of state (EOS). 
We plan to extend the code to include inelastic collisions, so as to investigate in future work how a realistic bulk viscosity ($\zeta/s\sim10^{-2}$) affects the system dynamics.

In order to extract the local thermodynamic quantities $T(x)$ and $\Gamma(x)$, one has to compute the energy density $e(x)$ and particle density $n(x)$ in the Local Rest Frame (LRF), which we choose to be the Landau LRF. We determine $e$, $n$, and the fluid four-velocity $u_{\mu}$ via:
\begin{equation}\label{eq:Landau_frame}
    T^{\mu\nu} u_{\nu} = e\, u^{\mu}, \quad n = n^\mu u_\mu
\end{equation}
where $T^{\mu\nu}$ is energy-momentum tensor and $n^\mu$ the four-density. 

We solve Eq. \eqref{eq:RBT} numerically by discretising the space-time in cells and using the test particle method, and sample $f(x,p)$ by considering a finite number $N$ of point-like test particles.
It can be shown that this method provides a solution of the Boltzmann equation given that the test particle coordinates satisfy the relativistic Hamilton equations of motion, which are solved with a 4th-order Runge-Kutta integration method on the time grid.
The collision integral is implemented through the stochastic algorithm as in Ref. \cite{Xu:2004mz,Ferini:2008he}. For more details on this implementation of the code see \cite{Nugara:2023eku, Nugara:2024net}.\\
The results reported here were obtained with a number of test particles of order 15M-20M. For each simulation (single curve) we run 30-60 events, until convergence is reached. Each space cell has $\Delta\eta_s=0.16$ and $\Delta x=\Delta y=0.2$~fm, apart from smaller systems (peripheral O-O and Ne-Ne) for which  $\Delta x=\Delta y=0.1$~fm. The code is written in C and has been optimised for High Performance Computing (HPC). 

\subsection*{Initial conditions}

The initial conditions of the Boltzmann equation  are specified by the distribution function $f_0(x,p)$ at $\tau=\tau_0$:
\begin{multline}
    f_0(x,p) =\gamma_0 \exp(-x_\perp^2/R^2)\,\theta(|\eta_s|- 2.5)\\\times \delta(\eta_s-Y)\exp{ \left(- \sqrt{p_x^2 + p_y^2 + m^2}/\Lambda_0 \right)} \label{eq:f_0}
\end{multline}
The initial distribution is uniform in space-time rapidity $\eta_s$ in the range $[-2.5,2.5]$, and Gaussian in the transverse plane ($\mathbf{x}_\perp = (x,y)$). The transverse radius $R$ determines the initial transverse extension of the fireball. In momentum space, we choose an initial distribution with $Y=\eta_s$, where $Y=\tanh(p_z/E)$ is the momentum rapidity. We fix $\Lambda_0$ and $\gamma_0$ to fulfil the matching conditions with the energy and particle density of a medium at equilibrium with temperature $T_0$ and fugacity $\Gamma_0$.

Following Ref.~\cite{Plumari:2015sia}, we induce spatial eccentricities $\varepsilon_n$  by deforming the initial
distribution, thus mimicking anisotropies due to the initial elliptic shape and/or initial state fluctuations~\cite{Alver:2010gr, Plumari:2015cfa}. To achieve this, we shift the particle coordinates by a small amount. Introducing the complex notation
$z=x+iy$, we shift $z$ according to 
\begin{equation}
\label{deformation}
z\to z+\delta z\equiv z-\alpha \bar z^{n-1},
\end{equation}
where 
$\bar z\equiv x-iy$, and $\alpha$ is a small real positive quantity chosen to get the desired eccentricity. In the specific case of a Gaussian initial distribution,
\begin{equation*}
    \varepsilon_n = \frac{n\,\Gamma(n)}{ \Gamma(1+n/2)} \alpha R^{n-2},
\end{equation*}
where $\Gamma(n)$ is the Euler Gamma function.

\begin{table}
\caption{Parameters for the smooth initial conditions. }
    \centering
    \begin{tabular}{ccccc}
        \toprule
        $\KnR$ & $R$ [fm] & $4\pi\eta/s$ & $\tau_0$ [fm] & $T_0$ [GeV] \\
        \midrule
        1.30 & 5.0 & 12.57 & 0.5 & 0.50 \\
        1.30 & 3.0 & 5.03 & 0.3 & 0.33 \\
        \midrule
        3.20 & 5.0 & 5.03 & 0.5 & 0.50 \\
        3.20 & 3.0 & 2.012 & 0.3 & 0.33 \\
        \midrule
        6.45 & 5.0 & 2.515 & 0.5 & 0.50 \\
        6.45 & 3.0 & 1.006 & 0.3 & 0.33 \\
        \bottomrule
    \end{tabular}
    \label{tab:parameters}
\end{table}

\section{Knudsen Number}
According to the literature, the Knudsen Number Kn is commonly defined as the ratio between a microscopic length, especially the mean free path $\mfp$, and a macroscopic scale. There is a certain degree of arbitrariness in choosing this macroscopic scale: following what was done in previous studies~\cite{Kurkela:2019kip,Ambrus:2021fej} we adopt the root-mean-square radius $R$ of the initial density distribution. One may argue that the time dependence of this transverse radius plays a role, since the system is expanding. However, we verified that our conclusions are not qualitatively affected by replacing the initial $R$ with a time-dependent $R(t)$. Accordingly we define the local inverse Knudsen number as:
\begin{equation}
\IKn(x) = \frac{R}{\mfp(x)}.
\end{equation}
Since we work on a discretised grid, $\mfp(x)\to\mfp^j$, with $j$ being the cell index. We define the local mean free path as:
\begin{align*}
\lambda_{\text{mfp}}^j &=  \frac{1}{N_{\rm test}^j} \sum_i^{N^j_{\rm test}}  \lambda_{\text{mfp}}^i = \frac{1}{N_{\rm test}^j} \sum_i^{N^j_{\rm test}} v_i\,t_i \approx \\
&\approx {t_{\rm coll}}\, \langle v_j\rangle =   \frac {\Delta\tau N_{\rm test}^j }{ 2 \,N_{\rm coll}^j}  \langle v_j\rangle
\end{align*}
where $N_{\rm test}^j$ is the number of test particles inside the $j$-th cell, $N_{\rm coll}^j$ is the number of collisions occurring in the given time-step, $\langle v_j\rangle$
is the average velocity and $t_i$ is the time interval between two successive collisions for each particle. Notice that we assume $t_i=t_{\rm coll}$ for every test particle according to the hypothesis of homogeneity within each cell, with $t_{\rm coll}$  being the average collision time per particle. In the massless case $v_i=1$, and therefore $\lambda_{\text{mfp}}=t_{\rm coll}$ exactly.

In order to have a unique quantity to describe the whole system, one has to compute a ``global'' $\IKn$. We can define an average weighted by the number of particles: 
\begin{align*}
    \langle\text{Kn}^{-1}\rangle &= 
    \frac{1}{\sum_j N_{\rm test}^j} \sum_{j}^{N_{\rm cell}} \frac{R}{\mfp^j} N^j_{\rm test} \\
   &=\frac{1}{\sum_j N_{\rm test}^j} \frac{2R}{\Delta \tau} \sum_j^{N_{\rm cell}} \frac{N^j_{\rm coll}}{\langle v_j \rangle},
\end{align*}
where the summation runs over the cells at midrapidity.
The role of the Knudsen number is strictly related to the regime of applicability of hydrodynamics. Indeed, as already stated in the Introduction, fluid dynamics is rigorously applicable if the microscopic ($\mfp$) and macroscopic ($R$) scales of the system are well separated, that is $\IKn\gg 1$. This means that we can understand, by looking at $\IKn(t)$, whether or not the evolution of the system could be modelled by fluid dynamics in its different stages.

As already outlined in Ref.~\cite{Nugara:2024net}, when solving the full kinetic theory at fixed initial geometry, the evolution of $\IKn(t/R)$ in conformal systems is universal: that is, if two systems share the same $\IKn$ at a certain $t/R$, they will share it throughout the whole evolution. This allows to cluster different collision systems in universality classes according to $\IKn$.
To label each of these classes, we choose the value of the inverse Knudsen number $\IKn (t=R)\equiv\KnR$, which has been shown to be similar in the conformal case to the opacity $\hat\gamma$ introduced in the context of RTA or ITA kinetic theory~\cite{Kurkela:2019kip, Ambrus:2021fej}.
A key point of the present work is that, while in RTA $\hat\gamma$ is the fundamental scaling parameter at least in the 2+1D conformal case, it is not the case in the context of the full relativistic Boltzmann equation. Indeed, in Ref.~\cite{Nugara:2024net} the scaling with opacity was shown to be broken sizeably, at least for exclusive observables like the ratio $v_n/\varepsilon_n$. As discussed below, instead, for both conformal and non-conformal systems the dynamical evolution and observables scale with the Knudsen number, which therefore appears as the key scaling parameter of the theory.

In particular, we choose two different systems with fixed $R, T_0, \tau_0$ and change $\eta/s$ to explore three values of $\KnR$, see Table~\ref{tab:parameters} for the detailed setup of the simulations. We refer to Sec.~\ref{sec:fluctuations} and in particular to Table~\ref{tab:event_by_event} to relate physical systems to universality classes, ranging from O-O to Pb-Pb at LHC and RHIC energies.

\begin{figure}
    \centering
    \includegraphics[width=1\linewidth]{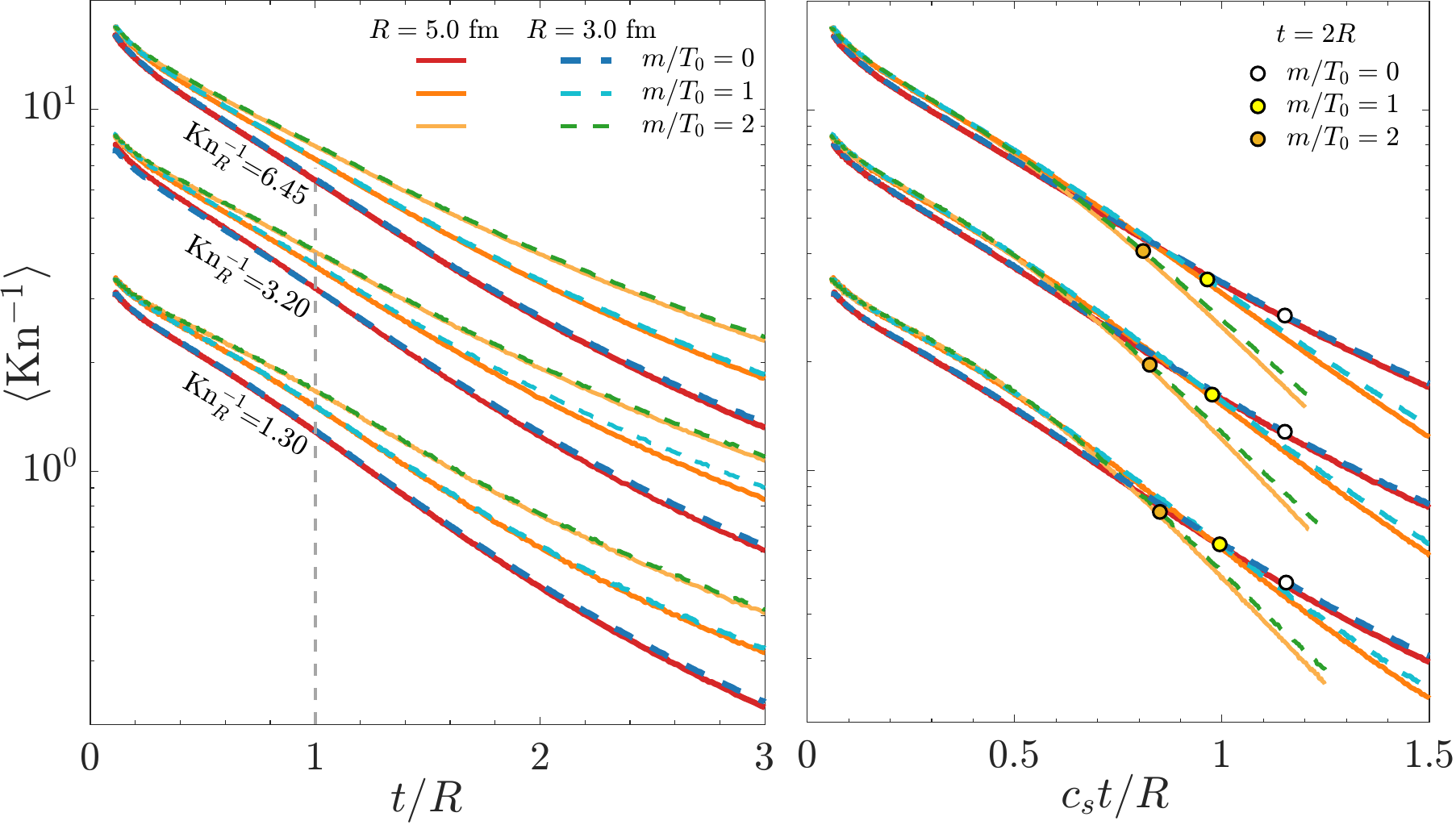}
    \caption{Dependence of the inverse Knudsen number on different scaled time variables (left: $t/R$; right: $c_s t/R$) for three different universality classes $\KnR=[1.30, 3.20, 6.45]$ and three different ratios $m/T_0=[0,1,2]$. The circles in the right panel correspond to $t/R=2$, after which scaling breaks down.}
    \label{fig:fig_1}
\end{figure}

In the non-conformal simulations, we keep the same initial conditions as in the conformal case and only change the mass. One clearly sees (left panel of Fig.~\ref{fig:fig_1}) that changing the ratio $m/T_0$ modifies significantly the functional form of $\IKn(t/R)$, with a subsequent loss of universality. By looking at the Figure, the comparison between red solid and green dashed lines shows that a scaling behaviour persists only if the ratio $m/T_0$ is kept fixed.

This observation suggests that a major role could be played by the speed of sound $c_s$: in the non-conformal case, indeed, $c_s=c_s(m/T)$, which means that perturbations travel through the medium with a velocity depending on the ratio $m/T$.
The typical macroscopic time scale is the time necessary for a perturbation to travel across the whole system, that is $R/c_s$.
In the conformal case there is obviously no difference between different systems since $c_s^2=1/3$ without any temperature dependence, which allows to recover the typical scaling variable $t/R$;
for a non conformal system, instead, one should replace it with $c_s t/R$. As shown in the right panel of Fig.~\ref{fig:fig_1}, with this choice of the scaled time it becomes possible to recover a rather good scaling, which only breaks down when $t>2R$ (see coloured circles), i.e. when the system is largely decoupled. Note that the speed of sound is a decreasing function of $m/T$.

Following Ref.~\cite{Ambrus:2023qcl}, for a particle-number conserving system of massive particles one gets:
\begin{align}\label{eq:sound_velocity}
    c_s^2 &= \left( \frac{\partial P}{\partial e} \right)_n + \frac{1}{h} \left( \frac{\partial P}{\partial n} \right)_e=\\
    &=\frac{4+z^2- g^2(z) + 3g(z)}{(3+z^2- g^2(z) + 3g(z))(g(z) +1)},
\end{align}
 where $g(z)$ is the ratio $e/P$ and $h$ the enthalpy per particle:
\begin{equation*}
    g(z)=\frac{e}{P} = z\frac{K_3(z)}{K_2(z)} -1, \quad h = \frac{e+P}{n} = m\frac{K_3(z)}{K_2(z)};
\end{equation*}
$K_n(z)$ are the modified Bessel functions of order $n$.
In the systems under examination the mass is fixed, but the temperature depends on space-time coordinates. By assuming local thermal equilibrium, we can compute the local $c_s(x)=c_s(m/T(x))$. In order to have a ``global'' sound velocity we define
$c_s=c_s(m/T_{\text{avg}})$, where $T_{\text{avg}}$ is the weighted average of the temperature. We checked that the temperature average can be weighted by energy density or particle density, with a difference $<1\%$ between the two.

\section{Collective flows}

\begin{figure}
    \centering
    \includegraphics[width=1\linewidth]{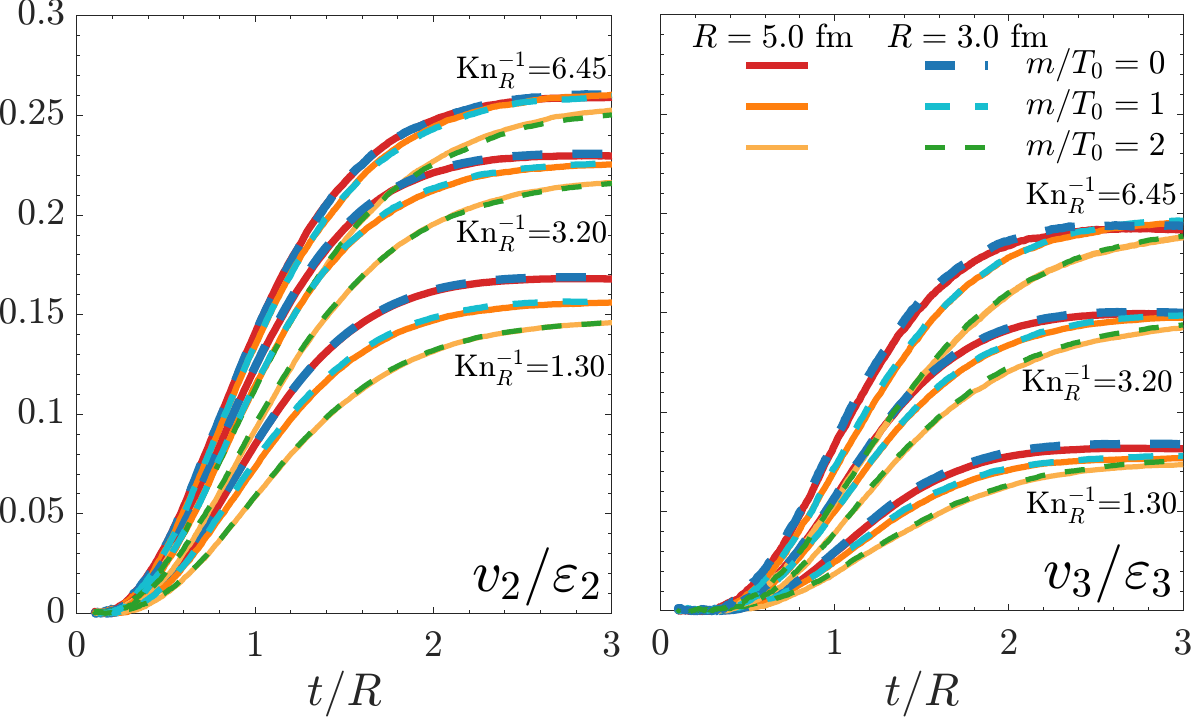}\vspace{8pt}
    \includegraphics[width=1\linewidth]{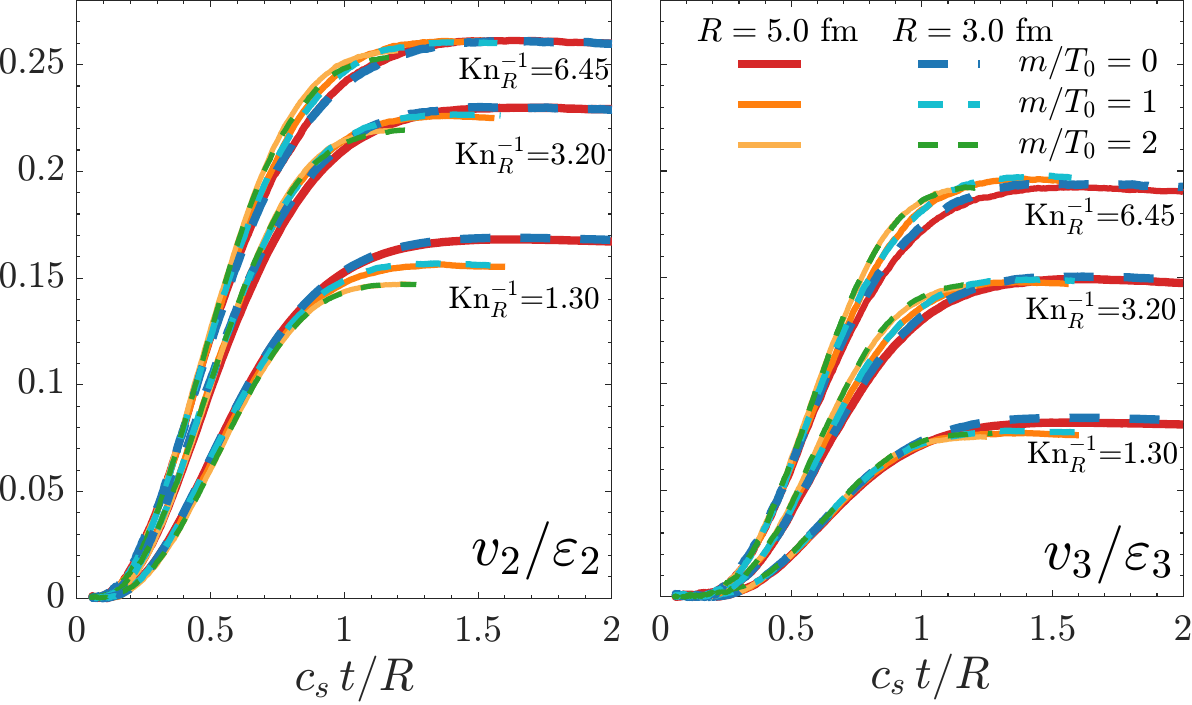}
    \caption{Response curves for elliptic (left panels) and triangular flows (right panels) in systems with fixed $\IKn$ and radii $R=5.0$ fm (red and orange curves) and $R=3.0$ fm (blue and green curves) (see Table \ref{tab:parameters} for the details) for different values of $m/T_0=[0,1,2]$. The top panels are plotted with respect to $t/R$, the bottom panels with respect to $c_st/R$.
    }
    \label{fig:fig_2}
\end{figure}

\begin{figure}
    \centering
    \includegraphics[width=1\linewidth]{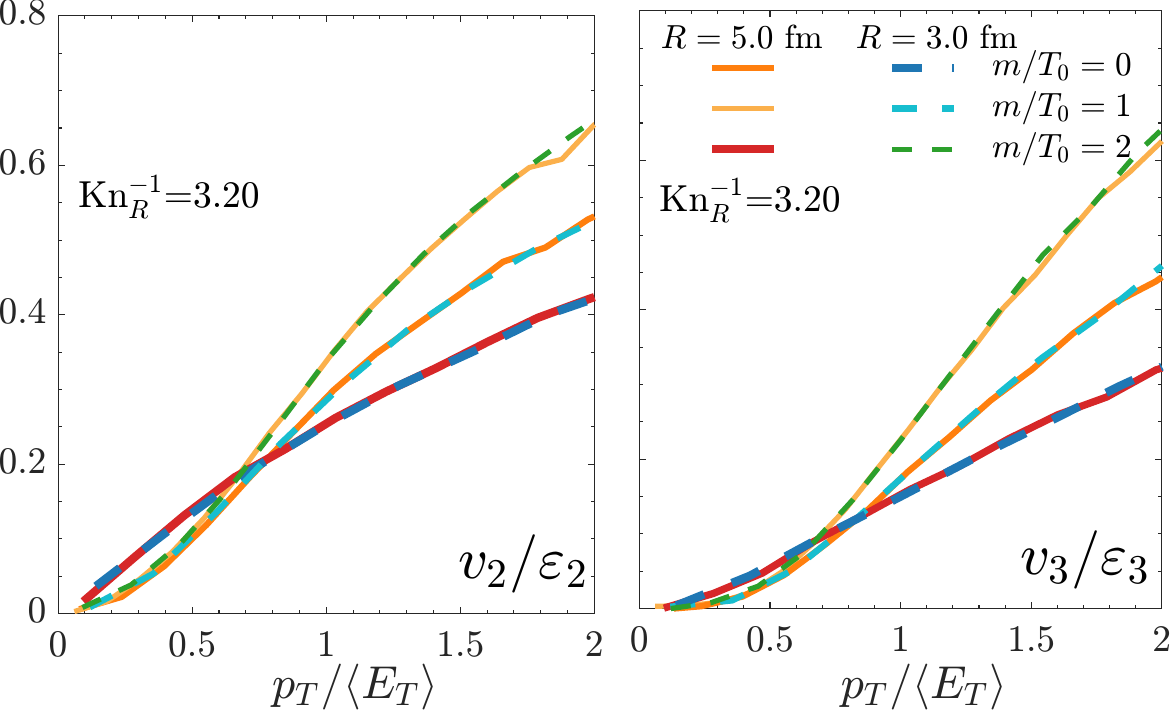}\\
    \caption{Differential elliptic (left) and triangular (right) flow at $t=3R$ plotted with respect to the scaled transverse momentum $p_T/\langle E_T\rangle$. The colour map is the same as Fig.~\ref{fig:fig_2}.}
    \label{fig:fig_3}
\end{figure}

In this section we investigate whether the response curves $v_n/\varepsilon_n$ ($n=2,3$) exhibit a universal scaling at fixed Knudsen number in the non-conformal case, as was been found in the conformal limit.
Note that all curves shown in Figs.~\ref{fig:fig_2} and \ref{fig:fig_3} have been obtained with $\varepsilon_n=0.2$, but that we checked that the results are independent of $\varepsilon_n$ in $[0.05,0.4]$, in which range it is known that both hydrodynamics and transport approaches show a linear dependence of the $v_n$ on $\varepsilon_n$~\cite{Noronha-Hostler:2015dbi, Roch:2020zdl, Plumari:2015cfa}.

In the top panels of Fig.~\ref{fig:fig_2} we show the scaled-time evolution of elliptic (left panel) and triangular (right panel) flow for three different $\IKn_R$ (see Table \ref{tab:parameters}) and three values of $m/T_0=[0,1,2]$. As a first check, we recover the familiar results of the conformal case: the thick dark red lines ($m=0$, $R=5.0$ fm) and the thick dashed dark green lines ($m=0$, $R=3.0$ fm) perfectly overlap for both $v_2$ and $v_3$ over the range of explored $\IKn$.

In the non-conformal case, the response functions show a clear a mass ordering, with larger masses leading to a smaller $v_n/\varepsilon_n$. This can be traced back to the fact that at larger $m/T$ the speed of sound is smaller, leading to a slower conversion from $\varepsilon_n$ to $v_n$; then the finite size of the system makes this conversion less efficient by dumping the process at $t>2R$, when the system is about to decouple. This effect is more evident in the range of small $\IKn$, where the system exhibits a particle-like behaviour, while in the hydrodynamic limit (large $\IKn$) the splitting between different $m/T_0$ is far less pronounced. In fact, the relative splitting at $\KnR=6.45$ for $v_2$ is $\lesssim3\%$, while at $\KnR=1.3$ it is $\approx 15\%$; similar deviations are observed for $v_3$.

To corroborate this interpretation, we plot in the bottom panel of Fig.~\ref{fig:fig_2} the response curves as function of $c_s t/R$. This has been shown to be the scaling variable for the collective flows also in the ideal-fluid limit \cite{Bhalerao:2005mm}.  One clearly sees that, by taking into account the role of the speed of the sound in the conversion $\varepsilon_n\to v_n$, it is possible to nearly recover the universality in the response functions. Note that, in agreement with Fig.~\ref{fig:fig_1}, the departure from universality occurs at $t\approx 2R$.

By moving to more differential observables, it is possible to become more sensitive to the microscopic details of the system, for example studying the differential flows $v_n(p_T)$. However,  one can expect to recover the universality by fixing the $\IKn$. In such observables the only relevant scale for transverse momentum $p_T$ for a single event is the average transverse energy $\langle E_T\rangle= \langle (p_T^2 + m^2)^{1/2} \rangle$. Similarly to what has been observed for the spectra in Ref.~\cite{Muncinelli:2024izj}, we study $v_n(p_T/\langle E_T\rangle)$: in Fig.~\ref{fig:fig_3} we display the flow values for $\KnR=3.20$ at $t=3R$, where the $v_n/\varepsilon_n$ have reached their saturation value. Interestingly, different systems (dashed curves: $R=3.0$ fm; solid curves: $R=5.0$ fm) with the same $m/T_0$ show a perfect scaling, pointing out that the universality seen in integrated observables (e.g. Fig.~\ref{fig:fig_2}) is still present even for differential quantities. Since in this Letter we focus on the bulk of the system, we show $v_n(p_T/\langle E_T\rangle)$ in the range $p_T/\langle E_T\rangle<2$, which includes more than $90\%$ of the spectrum. We note that this observable is strongly sensitive to the ratio $m/T_0$, which translates into a direct dependence on the speed of sound. In particular, at $p_T<\langle E_T\rangle$ we observe the typical linear vs quadratic increase of $v_2$ for the massless vs massive bulk~\cite{Danielewicz:1994nb}: the larger the mass, the slower the small $p_T$ dependence. 
On the other hand, for $p_T\gtrsim\langle E_T\rangle$ we observe an inversion of the mass-ordering. 
This can be partly explained by the fact that the values of $\langle E_T\rangle$ are larger for larger $m/T_0$. Thus, the simulations with $R=5.0$~fm, for instance, yield $\langle E_T\rangle=[1.1, 1.3, 1.6]$~GeV respectively for $m/T_0=[0, 1, 2]$. Nevertheless, this cannot fully explain the inverse mass ordering at large $p_T$. 
In more detailed investigations we found that it also follows from the non-physical fixing of $\eta/s$ also for $t>2R$ and $r>2R$, when the system is mostly decoupled and has already cooled down ($T<T_c$ pseudo-critical temperature): at small temperatures the local cross-section has to be much larger in the massive case in order to keep $\eta/s$ constant \cite{Plumari:2012ep}, which results in a larger response. This high-$p_T$ effect disappears  with a more realistic $\eta/s(T)$, which naturally implements a freeze-out for $T<T_c$. Indeed, as shown also in Ref. \cite{Plumari:2019gwq}, in this case the $v_n(p_T)$ exhibit the expected mass-ordering in the $p_T$ region explored.

\section{Impact of initial fluctuations}
\label{sec:fluctuations}

\begin{table}
\caption{Colliding systems studied in event-by-event simulations.}
    \centering
    \begin{tabular}{ccccc}
        \toprule
        system & energy  & centrality class  & $\IKn_R$ & $\langle R\rangle$ [fm] \\
        \midrule
         Pb-Pb & 2.76 TeV & 10--20\% &  6.5 & 4.20 \\
         Pb-Pb & 2.76 TeV & 40--50\% &  4.2 & 3.25 \\
         Pb-Pb & 2.76 TeV  & 60--70\%  & 3.0 & 2.72 \\
         \midrule
         Au-Au& 200 GeV & 20--30\% & 4.3  & 3.73 \\
         Au-Au& 200 GeV & 40--50\% & 3.2  & 3.20 \\
         \midrule
         Ne-Ne (1) & 7 TeV & 0--10\% & 4.1 & 2.75 \\
         Ne-Ne (1) & 7 TeV & 20--30\% & 3.2 & 2.5 \\
         Ne-Ne (2) & 7 TeV & 60--70\% & 1.8  & 1.92 \\
         \midrule         
         O-O (1) & 7 TeV & 20--30\% & 3.0  & 2.30 \\
         O-O (1) & 7 TeV & 60--80\% & 1.8 & 1.80 \\
         \bottomrule
    \end{tabular}
    \label{tab:event_by_event}
\end{table}

In order to have a more realistic description of the initial state of the collisions, we employ the well-known model \trento\:\cite{Moreland:2014oya} to generate initial profiles for different collisions systems (see Table \ref{tab:event_by_event}). Following the Bayesian analysis posterior performed for viscous hydro simulations \cite{Liyanage:2023nds}, we fix the \trento\ parameters $w = 0.985$ and
$p = 0.038$, which represent respectively the width of the Gaussian modeling the nucleon density profile and the reduced-thickness parameter that regulates the energy deposition of the participant nucleons. In order to identify the centrality classes, we generate $5\times 10^5$ events and classify them in ten classes of ten percentiles each (0-10\%, 10-20\%...), by making use of the matching formula between initial energy density and final charged multiplicity  at fixed $\eta/s$ introduced in Ref.~\cite{Giacalone:2019ldn}, $dN_{\rm ch}/d\eta \propto \int d^2x \,[e(x)\tau]_0^{2/3}$, as done e.g.\ in Ref.~\cite{Andronic:2025ylc}. The generated outputs are interpreted as energy-density profiles and used to map the initial distribution function. In order to construct the full $f(x,p)$, we have to assume an equation of state and an initial momentum distribution: as a starting point we consider a conformal system ($m=0$) with an effective number of degrees of freedom $n_{\text{dof}}=49$ and an initial momentum distribution with $Y=\eta_s$ (see Section \ref{sec:RBT}). We plan to study non-conformal event-by-event simulations in future work~\cite{Nugara@2025}. Moreover, since we only have a 2D profile from \trento, we assume the system is boost-invariant in the space-time rapidity interval $\eta_s\in[-2.5, 2.5]$. The only free parameter we need to fix in the matching \trento+RBT is the overall scaling factor. We fix it by comparing the final $(dE_T/d\eta_s)_{\tau_\text{freeze}}$ extracted from the code and the experimental  data or predictions.
The freeze-out surface, on which $(dE_T/d\eta_s)_{\tau_\text{freeze}}$ is computed, is defined by a condition on the energy density: when one cell is below the threshold $e<0.182$ GeV/fm$^3$, its energy is added to $(dE_T/d\eta_s)_{\tau_\text{freeze}}$. The freeze-out energy-density corresponds to the value $e(T_c)$ assumed at the critical temperature according to the lattice QCD EOS \cite{HotQCD:2014kol, Borsanyi:2010bp}. Finally, we fix $\eta/s=0.13$ for all these simulations, in agreement with studies performed for LHC energies with relativistic viscous hydrodynamics and transport theory \cite{Gale:2012rq, Romatschke:2007mq, Song:2008si, Xu:2004mz, Ferini:2008he, Konchakovski:2012yg}. In particular, for Pb-Pb collisions at $\sqrt{s_{\rm NN}}={2.76}$~TeV we use the experimental data for $dE/d\eta$ at midrapidity from ALICE~\cite{ALICE:2016igk}; for Au-Au at $\sqrt{s_{\rm NN}}={200}$ GeV we use PHENIX results~\cite{PHENIX:2013ehw}. For O-O and Ne-Ne collisions at LHC energy, since we miss experimental data, we fix it by following existing predictions and using the rough estimate $dE_T/d\eta_s\approx (3/2)\langle p_T\rangle\, dN_{\rm ch}/d\eta_s$. Since there is not agreement among the different predictions, we label by (1) the simulations gauged on the predictions by Refs.~\cite{Khan:2024fef, Behera:2021zhi} and by (2) those following Ref.~\cite{Giacalone:2024luz}. Moreover, to take into account the peculiar nucleon configurations of $^{16}$O and $^{20}$Ne, we use the NLEFT configuration files with positive weights from Refs.~\cite{Summerfield:2021oex, Giacalone:2024luz}.

\begin{figure}
    \centering
    \includegraphics[width=1\linewidth]{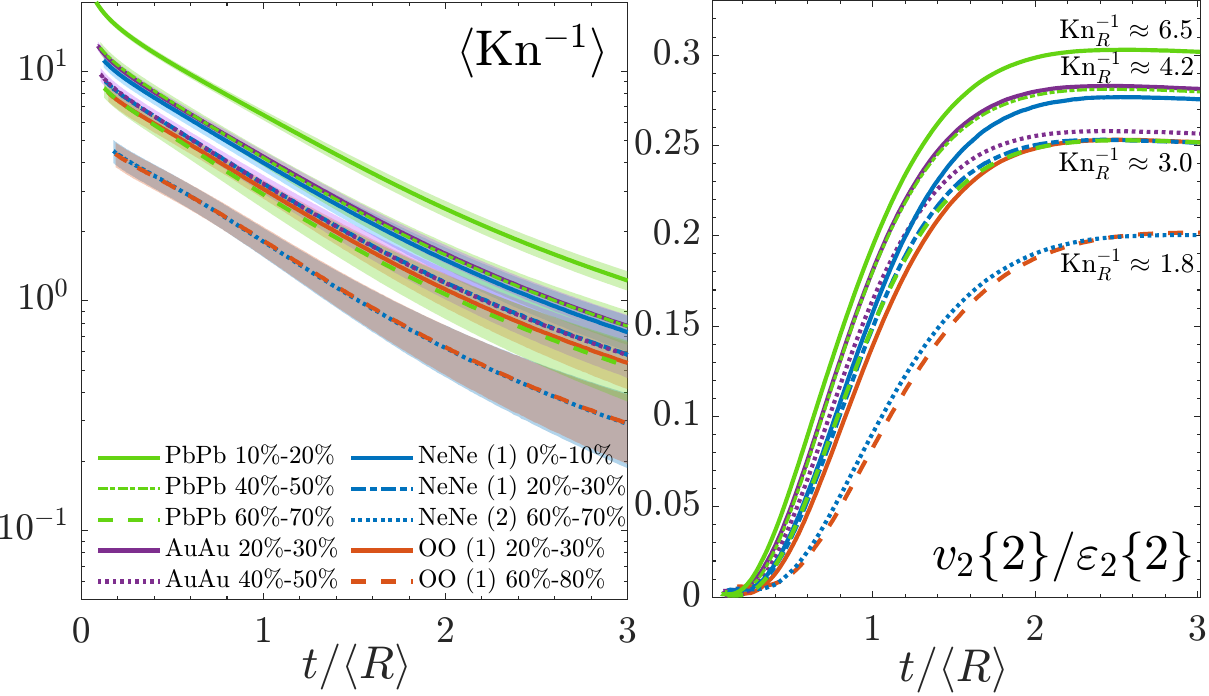}
    \caption{Left panel: Inverse Knudsen number evolution for different collision systems, energies and centrality classes. From the top to the bottom, they correspond to a $\IKn_R\approx6.5, 4.2, 3.0, 1.6$. More details on the event-by-event configurations in Table \ref{tab:event_by_event}. Right panel: Response functions for the same simulations of left panel. It is evident the scaling behaviour irrespectively of energy, system size and centrality.}
    \label{fig:fig_4}
\end{figure}

Following the discussion of the previous sections, we want to study whether, also in the more realistic case with initial state fluctuations, one can cluster different events in the same universality class. In order to compare with what has been studied above, we choose centrality classes in order to get an average inverse Knudsen number close to the ones discussed above (6.45, 3.20, 1.30), which allow us to explore a wide range of system sizes from central Pb-Pb to peripheral O-O and Ne-Ne; we also added a fourth value ($\KnR \simeq 4.2$) to probe further configurations, such as semi-central Au-Au and central Ne-Ne. The results shown below are obtained with 200 events for each ensemble. Obviously, the time evolution of $\IKn$ differs for each event of a given class: in the left panel of Fig.~\ref{fig:fig_4} we show the evolution of the average inverse Knudsen number and a band representing one standard deviation of the fluctuations occurring across the event-by-event simulations. The $R$ used here is the average of the root-mean-square radii of the different events. In Table \ref{tab:event_by_event} we list and give details about these simulations.

Regarding collective flow, we show in the right panel of Fig.~\ref{fig:fig_4} the two-particle correlation elliptic flow \cite{Borghini:2000sa, Borghini:2001vi} $$v_2\{2\} = ( c_{v_2}\{2\})^{1/2} \approx (\langle v_2^2\rangle)^{1/2},$$
scaled by the two-particle correlation  $\varepsilon_2\{2\}$, defined in an equivalent way. Here $\langle\cdot\rangle$ indicates an average over the ensembles.
Comparing with Fig.~\ref{fig:fig_2}, we observe systematically larger responses in the event-by-event simulations with respect to the ones with smooth initial conditions. This is not surprising since the detailed geometry of the energy and particle density distribution has to play a major role in building the anisotropic flow. By looking at the space-time development of elliptic, we found that this difference is mainly due to the presence of strong temperature gradients (``hot spots'', which are absent in the smooth initial conditions defined by Eq.~\ref{eq:f_0}, with a uniform $T_0$). Indeed, when particles exchange momenta in a region with higher $T$, they have on average a larger $p_T$ than the surrounding area, therefore they can easily cross it and carry the generated flow out of the hot spot, producing a larger response. On the contrary, if no such peaks are present, particles continue colliding and exchanging momenta until they reach the edges of the distribution, making less efficient the building of the collective flow. We will discuss more in detail the space-time development of the $v_n$ in Ref.~\cite{Nugara@2025}.

Looking more in detail to the left panel of Fig. \ref{fig:fig_4} and to Table \ref{tab:event_by_event}, one sees that, e.g., 60--70\% Pb-Pb, 40--50\% Au-Au, 20--30\% Ne-Ne and 20--30\% O-O collisions exhibit a very similar evolution of $\IKn$, with $\KnR\approx 3$ and average radii $\langle R\rangle = 2.3$--3.25 fm. In the left panel of the figure, the corresponding $v_2\{2\}/\varepsilon_2\{2\}$ curves evolve towards a very similar saturation value. The same can be said also for the other sets of simulations sharing the $\IKn$ evolution: 40--50\% Pb-Pb, 20--30\% Au-Au and 0--10\% Ne-Ne at $\KnR\approx4.2$; 60--70\% Ne-Ne and 60--80\% O-O at $\KnR\approx1.8$. This is particularly interesting since it is non-trivial that the scaling behaviour of the response functions observed within the context of events with smooth initial conditions, all sharing the same geometry, is now recovered among event-by-event simulations. The geometry of each initial profile is different, with the likely formation of peaks and strong anisotropies, especially for small systems and for the highly deformed $^{20}$Ne nucleus.  Nonetheless, when one averages on an ensemble of events, universality in the response function is fully recovered across systems with different nuclei and collision energies, ranging from $^{16}$O to $^{208}$Pb, from 200 GeV (Au-Au collision at RHIC) to 7 TeV. Notice that this is true also for values of the inverse Knudsen number ($\KnR\approx 2$--3) which are certainly out of the hydrodynamic regime. 

We emphasise once again that the only parameter we have the freedom to choose is $\eta/s$, which however is fixed for all the simulations. It would be of great interest to study how a more realistic $T$-dependent $\eta/s$ and a non-conformal equation of state affect these results. We will address this investigation in Ref.~\cite{Nugara@2025}. 

\section{Conclusions}
In this Letter, we investigated the role of the Knudsen number as the key scaling parameter characterising the dynamical evolution of QGP matter produced in uRHICs. We have employed the 3+1D Relativistic Boltzmann Transport approach and explored a wide range of initial conditions, corresponding to different collision systems. We moved from the full hydrodynamic regime ($\KnR=6.45$) to particle-like systems ($\KnR=1.30$), with a focus on the transition regime ($\KnR=3.20$).  For the first time, we extended the model to study both conformal and non-conformal systems. The presence of universality classes in the Knudsen number evolution, previously recognised only in the massless case, is recovered in this broader context, provided the role of the speed of sound is taken into account. In particular, one has to rescale the time $t$ with respect to the typical time scale in which perturbations propagate through the medium, namely $R/c_s$. By doing so, universality is nearly completely recovered in the evolution of the inverse Knudsen number, even for quite different setups: the only relevant information is the class to which the system belongs, which identifies the hydro- or particle-like behaviour until decoupling sets off at $t\sim2R$. This suggests that a scaling driven by $\IKn$ may be visible also in observables, such as the response functions $v_n/\varepsilon_n$. This hypothesis is confirmed by the fact that, if one studies the $v_n/\varepsilon_n$ as function of the scaled time $c_s t/R$, the evolution of the build-up of the anisotropic flows depends primarily on $\IKn$; however, the saturation value of the $v_n$ is settled by the transverse size $R$, which defines the time at which the system is about to decouple.
Finally we studied a more realistic case with initial-state fluctuations and solving event-by-event the Relativistic Boltzmann Transport equation in the massless case, still with a fixed constant $\eta/s=0.13$. Also in this setup we find that the Knudsen number is the key parameter that controls the response function $v_2\{2\}/\varepsilon_2\{2\}$. Irrespective of the collision system (from O-O to Pb-Pb), collision energy (from 200 GeV to 7 TeV), centrality class, and system size (from average radius $\langle R\rangle=1.8$ fm to $\langle R\rangle=4.2$ fm), ensembles of events sharing the same average $\IKn$ show the same response. For instance, our study suggests that Pb-Pb 60--70\% at 2.76 TeV, O-O and Ne-Ne 20--30\% at 7~TeV (LHC) and Au-Au 40--50\% at 200 GeV (RHIC) generate a very similar $v_2\{2\}/\varepsilon_2\{2\}$ response; the same is true for the other explored sets of simulations (Pb-Pb 40--50\%, Au-Au 20--30\% and Ne-Ne 0--10\%; Ne-Ne 60--70\% and O-O 60--80\%), regardless of the $\KnR$ value and hence independent of whether the systems lie in the hydrodynamic or particle-like regime.

These findings suggest that, in the context of the full kinetic theory, the Knudsen Number acts as the fundamental scaling parameter, governing the dynamical evolution of the medium in different regimes and in a broad range of scenarios.

In the future, aiming to a fully realistic study, we will investigate whether this key role of $\IKn$ also holds for observables, including more differential ones, computed in collisions with fluctuating initial conditions and a system with a non-conformal equation of state, and whether it also persists in the case of a non-constant $\eta/s(T)$, especially at $T<T_c$ and when including inelastic processes in the evolution.

\section*{Acknowledgments}

We thank Fernando Gardim and S\"oren Schlichting for discussions.
V.N. thanks the High Energy Theory group at Bielefeld University for the kind hospitality, which enabled this work.
N.B. acknowledges support by the Deutsche Forschungsgemeinschaft (DFG, German Research Foundation) through the CRC-TR 211 'Strong-interaction matter under extreme conditions' - project number 315477589 - TRR 211.


\end{document}